\documentclass{article}
\usepackage{ijcai24}
\usepackage{times}
\usepackage{soul}
\usepackage{url}
\usepackage[hidelinks]{hyperref}
\usepackage[utf8]{inputenc}
\usepackage[small]{caption}
\usepackage{graphicx}
\usepackage{amsmath}
\usepackage{amsthm}
\usepackage{booktabs}
\usepackage{algorithm}
\usepackage{algorithmic}
\usepackage[switch]{lineno}
\usepackage{multirow}
\usepackage{amsmath}
\usepackage{xcolor}
\usepackage{colortbl}
\usepackage{amssymb}
\usepackage{hhline}

\newif\ifshowcomments
\showcommentstrue 
\ifshowcomments
    \newcommand{\yj}[1]{{\color{blue}[YJ: #1]}}
\else
    \newcommand{\yj}[1]{}
\fi

\title{
Secure and Efficient Watermarking for Latent Diffusion Models in Model Distribution Scenarios}

\author{
Liangqi Lei$^1$
\and
Keke Gai$^1$\and
Jing Yu$^{2}$\And
Liehuang Zhu$^1$\\
Qi Wu$^3$\\
\affiliations
$^1$Beijing Institute of Technology\\
$^2$School of Information Engineering, Minzu University of China.\\
$^3$School of Computer Science, The University of Adelaide. \\
\emails
3120245873@bit.edu.cn, gaikeke@bit.edu.cn,
jing.yu@muc.edu.cn,
liehuangz@bit.edu.cn,
qi.wu01@adelaide.edu.au
}

\begin{document}

\maketitle

\begin{abstract}

Latent diffusion models have exhibited considerable potential in generative tasks. Watermarking is considered to be an alternative to safeguard the copyright of generative models and prevent their misuse.
However, in the context of model distribution scenarios, the accessibility of models to large  scale of model users brings new challenges to the security, efficiency and robustness of existing watermark solutions. To address these issues, we propose a secure and efficient watermarking solution. A new security mechanism is designed to prevent watermark leakage and watermark escape, which considers  watermark randomness and watermark-model association as two constraints for  mandatory watermark injection. 
To reduce the time cost of training the security module, watermark injection and the security mechanism are decoupled, ensuring that fine-tuning VAE only accomplishes the security mechanism without the burden of learning watermark patterns. A watermark distribution-based verification strategy is proposed to enhance the robustness against diverse attacks in the model distribution scenarios.
Experimental results prove that our watermarking  consistently outperforms existing six baselines on effectiveness and robustness against ten image processing attacks and adversarial attacks, while enhancing security in the distribution scenarios.
The code is available at https://anonymous.4open.science/r/DistriMark-F11F/.
\end{abstract}

\begin{figure}[t]
\centering
\includegraphics[width=\columnwidth]{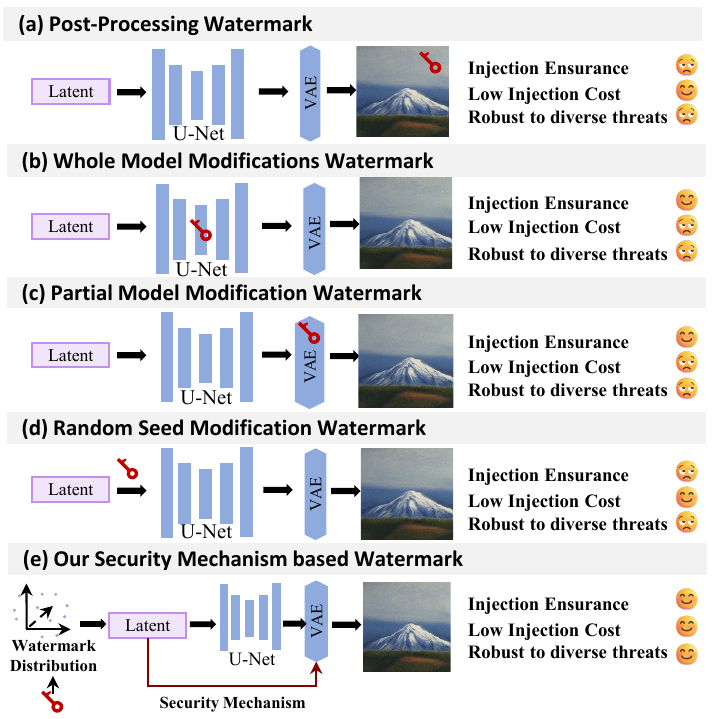} %
\caption{Watermark framework comparison with existing  solutions.}
\label{fig1}
\end{figure}

\section{Introduction}
Substantial progress in latent diffusion models (LDMs) \cite{croitoru2023diffusion} have significantly enhanced the quality of image generation, which presents observable abilities in producing a wide scope of creative visuals, e.g., artistic works and realistic depictions. 
To safeguard the copyright of generative models \cite{gowal2023identifying} and prevent their misuse \cite{barrett2023identifying}, watermarking is one avenue for detecting generated content and tracing its source. Recently, there is a compelling trend for  model producers  to distribute LDMs to numberous model users by model sharing \cite{donahue2021model}, disclosure \cite{azcoitia2022survey}, and trading \cite{pei2023data}. Since a large amount of model users are granted with model architecture access and fine-tuning permission in these model distribution scenarios, effective watermark injection and robust watermark verification becomes more challenging compared with local model usage. 

In order to support applications in  model distribution scenarios, LDM watermarks need to accommodate serveral key constrains. (1) Since model networks and parameters will be distributed for personalized usage, it is possible for model users to bypass the  watermark injection by model modifications. Therefore, security mechanism is indispensable to avoid watermark evading. (2) When the model is distributed to massive users, the watermark has to guarantee low injection time cost while spanning distinctive information for a large amount of user verification. (3) Due to the higher model access permission and larger user scale in model distribution scenarios, untrustworthy users pose a greater threat of model theft and leakage, making it essential for watermarking methods to ensure robustness against diverse adversaries.

A traditional watermarking solution is post-processing watermark that embeds watermarks after image generation (Figure \ref{fig1}(a)). However, untrustworthy users can remove post-hoc watermark trivially. 
On the other hand, in-processing watermarks inject messages into the image generation process, which contain three category solutions based on modification ways. Whole model modifications \cite{zhao2023recipe,feng2024aqualora} embed watermarks by training the entire generative models (Figure \ref{fig1}(b)), which require substantial training resources and thus inefficient in terms of model distribution senarios. Partial model modifications \cite{fernandez2023stable,xiong2023flexible} merely fine-tune the decoder of the LDMs (Figure \ref{fig1}(c)). However, these methods are vulnerable to multiple attacks \cite{an2024benchmarking} such as reconstructive attack \cite{zhao2023invisible} and adaptive adversarial sample attack \cite{jiang2023evading}. Random seed modifications \cite{wen2024tree,yang2024gaussian,ci2025ringid} 
inject watermarks into the initial latent variable of LDMs which are time-efficient without model fine-tuning and robust against diverse attacks. But in model distribution scenarios, the untrustworthy user can easily change the initial latent vector to circumvent watermark injection.

In this work as shown in Figure \ref{fig1}(d), we extend the application of LDM watermarking to model distribution scenarios and propose a secure and efficient watermarking method, named as DistriMark. Considering the watermark injection efficiency, DistriMark is based on the random seed modification schema without any model fine-tuning. To avoid model user bypassing watermark injection, we propose a watermark-network controller module as a security mechanism, which establishes binding association between the VAE network in LDMs and the watermarked initial latent variable. In this way, LDMs can generate expected content only when the watermark is mandatory injected. To reduce the time cost of training the watermark-network controller module, we decouple the watermark injection and the security mechanism, ensuring that fine-tuning VAE only accomplishes the security mechanism without the burden of learning watermark patterns. Furthermore, we propose watermark generation module to transform the watermark into a watermark-specific distribution and obtain a watermarked latent variable through sampling strategy. For watermark verification, the  latent variable obtained by diffusion inversion is compared to the watermark distribution instead of a fixed watermarked variable. 
This watermark generation and verification strategies not only increases the security of plaintext watermarks, but also makes up the errors caused by diffusion inversion and enhance the robustness against various watermark attacks.

\begin{figure*}[t]
\centering
\includegraphics[width=2\columnwidth]{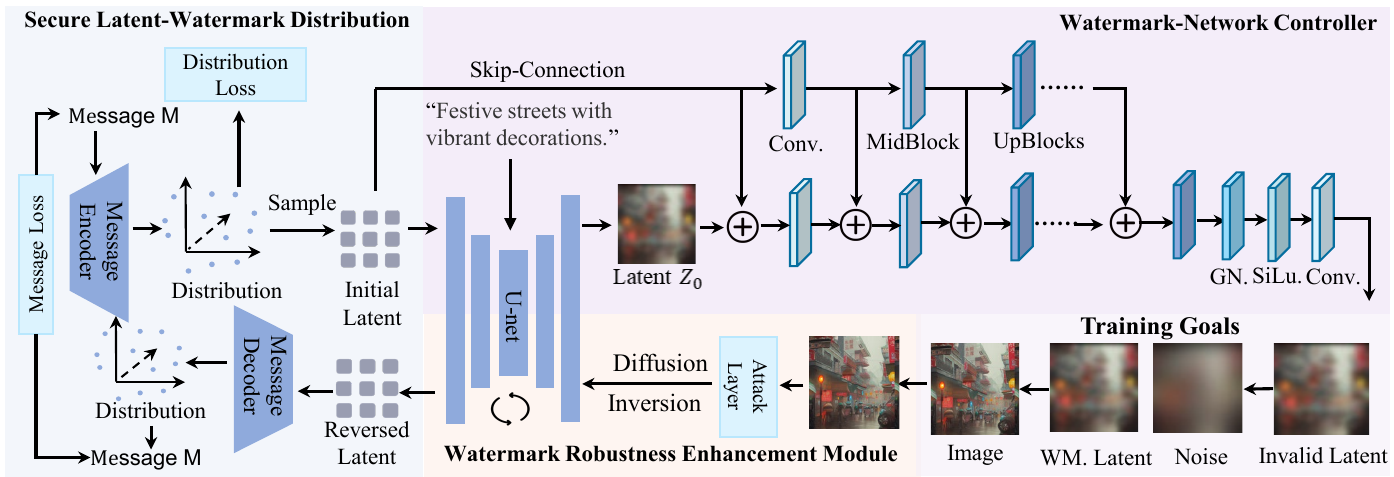} %
\caption{Framework of the proposed DistriMark watermarking scheme for Latent Diffusion Model.}
\label{fig2}
\end{figure*}

The main contributions are summarized as follows:
(1) We propose new security mechanism to prevent watermark leakage and watermark escape in the model distribution scenarios by pseudo-random latent variable transformation and VAE-based fine-tuning strategy. We consider watermark randomness and watermark-model association as two constraints for enhancing watermarking security, which sheds new light on the real-world application of diffusion model watermarking.
(2) We propose a novel model distribution scenario-oriented watermarking schema for LDMs. By injecting multi-bit watermarks into the initial latent variables and fixing the verification errors via watermark distribution verification and adversarial training strategy, our schema achieves both robustness and flexibility compared with existing fine-tuning and random seed-based watermarks.
(3) DistriMark shows superior performance on effectiveness and robustness compared with existing six baselines over ten image processing attacks, challenging adaptive adversarial sample attacks and reconstructive attacks. DistriMark is more secure against watermark escape and leakage compared with existing random seed modification watermarks in the distribution scenarios.


\section{Related Work}


\textbf{Diffusion Models} has demonstrated prominent performance in image generation \cite{dhariwal2021diffusion} with the support of methodologies \cite{song2020score} and sampling techniques \cite{song2020denoising}.
Latent diffusion models optimize images in the latent space of pre-trained VAEs, further accelerating the practical applications of diffusion models while also raising concerns about potential abuse and intellectual property of models. 
The immense cost of training a diffusion model, which requires hundreds of GPU-days \cite{rombach2022high}, makes copyright protection for diffusion models crucial, especially when the model architecture and weights are distributed to users for deployment. We focus on the security and efficiency issues of model watermarking in distribution scenarios.

\noindent\textbf{Watermarking for Latent Diffusion Models} is primarily aimed at tracing the origins of generated images of the latent diffusion model. WDM \cite{zhao2023recipe} trains an autoencoder to stamp a watermark on all training data before re-training the generator from scratch. However, this approach suffers from inefficiencies in terms of computational resources and time. Stable Signature \cite{fernandez2023stable} and FSwatermark \cite{xiong2023flexible} fine-tune VAE-Decoder to ensure that all generated images contain the watermark. However, these approaches are not resilient to diverse threats. Tree-ring \cite{wen2023tree} and ZoDiac \cite{zhang2024robust} propose random seed modification watermarks which show significant advantages in dealing with various processing attacks \cite{an2024benchmarking}. However, these methods lack secure mechanisms to guarantee watermark embedding in model distribution scenarios.

\noindent\textbf{Model Watermarking Attacks}
on diffusion model watermarking primarily occur at two levels: image and model. At the image level, attacks such as image processing attacks, adaptive adversarial sample attacks \cite{jiang2023evading}, and reconstruction attacks \cite{zhao2023invisible} are included. At the model level, attacks include techniques such as purification and model collison. 
Model purification will significantly reduce the detection accuracy of whole model modification watermark and partial model modification watermark. Model collison  will deceive watermark detection.
We propose watermark-network controller to avoid watermark verification issues related to model-level attacks and ensures image-level robustness by secure watermark distribution.
\section{Methodology}
\label{sec-3.2}

    

\subsection{Framework of DistriMark} 

In this work, we extend to achieve the security and embedding efficiency of watermarks in the model distribution scenarios. Our Distrimark embed the watermark into the latent variables of the diffusion model and enforce the mandatory embedding of the watermark whenever the model is utilized by leveraging the watermark-network controller.
To guarantee the security of watermark distribution and maintain the unpredictability and fidelity of watermark, we propose a novel watermark distribution method Secure Latent Watermark Distribution. This method establishes a unified representation of latent variables and watermark information as shown in Figure \ref{fig2}. The watermark region follows a specific distribution, from which watermarked latent variables are sampled. The variability in latent variables across different outputs increases randomness and unpredictability, which ensures the security of watermark distribution. To safeguard the security of watermark embedding, we introduce Watermark-Network Controller, a security mechanism integrated into latent diffusion model components which binds the variational autoencoder with watermarked latent variables to prevent users from evading the watermark embedding process. This module binds the VAE-Decoder with watermarked latent variables through skip connections. The image quality will significantly deteriorate when the model user escape the watermark. Distirimark utilizes a three-step progressive training strategy with the following objectives:

\subsection{Watermark-Network Controller}
To enforce the embedding of watermarks during model usage, the watermark-network controller directs image generation by using watermarked initial latent as control signals. watermark-network controller connects the watermarked initial latent variables and relevant components of the VAE-Decoder through skip connections. Through fine-tuning the VAE-Decoder, images corresponding to the watermarked latent variables consistent with the original model, while corresponding to random latent variables are transformed to random noise. In the implementation of skip-connection, we design a network association to bind the watermarked initial variable to the intermediate layer variables.

The loss function employs the LPIPS loss and L2 distance between images, denoted as $L_1$ and $L_2$, respectively. $\mathcal{L}_2(D_o(z),D_v(z))=||D_o(z)-D_v(z)||_2^2.$ $D(\cdot$) denotes the decoding process of the variational autoencoder $D_o$ and $D_v$ denote the original and the fine-tuned VAE-Decoder with skip connections respectively. When the VAE-Decoder is connected to the initial latent variable, the loss function is:
\begin{equation}
\mathcal{L}_w=L_1(D_o(z),D_v(z))+  L_2(D_o(z),D_v(z)))
\end{equation}

\noindent where $Z_r$ denotes the random noise. The factor $\lambda_v$ is a constant. When the VAE-Decoder is connected to the random latent variable, the loss function is:
\begin{equation}
\mathcal{L}_u=(L_2(D_o(z_{r}),D_v(z))-\lambda_v \times  L_2(D_o(z),D_v(z)))
\end{equation}

To prevent pixels with smaller values from being excessively altered, We calculate the difference across multiple channels between the output images of the original model and the fine-tuned model as the loss:
\begin{equation}
    \mathcal{L}_i=\frac{1}{c\times h\times w}\sum_{k=1}^c\sum_{i=1}^h\sum_{j=1}^w\frac{\left|D_v(z)_{(k,i,j)}-{D_o(z)}_{(k,i,j)}\right|}{D_v(z)_{(k,i,j)}+max(D_v(z))}
\end{equation}

\noindent where $\theta$ indicates whether the initial latent variables are from the message encoder. $\varepsilon,\delta$ is the balancing weight. The overall loss for this step is as follows:
\begin{equation}
\mathcal{L}_v=\theta\times\mathcal{L}_w+\varepsilon\times(1-\theta) \times \mathcal{L}_u+\delta\times \mathcal{L}_i
\end{equation}

\begin{figure*}[t]
\centering
\includegraphics[width=1.7\columnwidth]{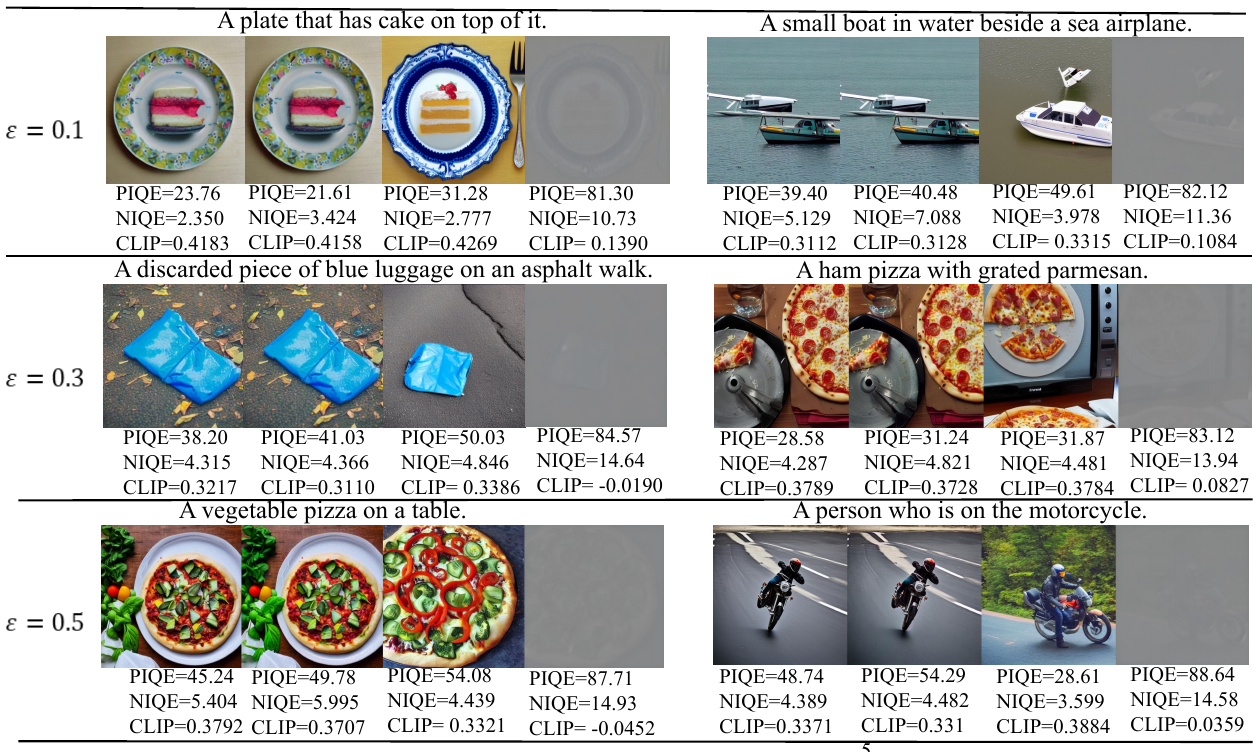} %
\caption{Generated image comparison under the security mechanism. Images in each sample from left to right are Watermarked Initial Latent Variables (WIL for short) without VAE-based fine-tuning (fine-tuning for short), WIL with fine-tuning, non-WIL without fine-tuning, and non-WIL with fine-tuning, representatively.}
\label{fig8}
\end{figure*}

\subsection{Secure Latent-Watermark Distribution}

We assume a series of deterministic functions $f(z;\theta)$ parameterized by a vector $\theta$. When $\theta$ is fixed and $z \sim \mathcal{N}(1,0)$, $f(z; \theta)$ can generate latent variables that conform to a specific distribution. Specially, the encoder outputs the mean vector and variance vector to simulate the deterministic function $f(z;\theta)$ and generates the initial latent variables through reparameterization. 

The watermarked latent variables are put into the message decoder directly to train them in the self-supervision paradigm. The message loss is the Binary Cross Entropy (BCE) between $m$ and the sigmoid $\sigma(m')$:



\begin{equation}
\begin{aligned}
    &\mathcal{L}_m
    =-\sum_{k=0}^{n-1}m_k\log\sigma(m_k')+(1-m_k)\log(1-\sigma(m_k'))
\end{aligned}
\label{step1_BCE}
\end{equation}

Since the training samples of the diffusion model are generated by progressively adding noise until they conform to a standard normal distribution, during the inference stage, the message encoder will output initial latent variables that follow the same distribution. The Kullback-Leibler (KL) divergence between the initial latent variables and the standard normal distribution is utilized as the loss function. The output follows a normal distribution, denoted as $q(z) \sim \mathcal{N}(\mu_1, \sigma_1^2)$ and the standard normal distribution is denoted as $p(z) \sim \mathcal{N}(0, 1)$. The distribution loss is as follows:

\begin{equation}
\begin{aligned}
    \mathcal{L}_d=D_{KL}(q(z)||p(z))
=\int_xq(x)log\frac{q(x)}{p(x)}dx
\end{aligned}
\end{equation}



\subsection{Watermark Robustness Enhancement Module}
\noindent\textbf{Watermarking Verification.} Watermark extraction involves diffusion inversion, an approximate process for obtaining initial hidden variables from generated images. 
Diffusion inversion \cite{dhariwal2021diffusion} algorithmically retrieves the initial latent variables from images generated by a diffusion model. $x_t$ represents the image at the timestep $t$. Based on the assumption $x_{t-1} - x_t 
\approx x_{t+1} - x_t$ , diffusion inversion of the Denoising Diffusion Implicit Model (DDIM) \cite{song2020denoising}  is formalized as follows:
\begin{equation}
\hat{x}_{t+1}=\sqrt{\bar{\alpha}_{t+1}} x_{0}+\sqrt{1-\bar{\alpha}_{t+1}} \epsilon_{\theta}(x_{t})
\end{equation}
where $\bar{\alpha}$ is the parameter of the diffusion model. $t$ denotes the denoising timestep.  $\epsilon_\theta(x_t)$ is the estimated noise for timestep $t$.
$\hat{x_{0}}$ represents the prediction of the image at the current timestep and is defined as:
\begin{equation}
    \hat{x_{0}}=\frac{x_{t}-\sqrt{1-\bar{\alpha}_{t}} \epsilon_{\theta}\left(x_{t}\right)}{\sqrt{\bar{\alpha}_{t}}}.
\end{equation}

To mitigate the effects of diffusion inversion and raise the robustness of image processing, we introduce the watermark robustness enhancement module which employs adversarial training to raise performance of the message decoder.
The loss function is binary cross entropy between the message $m$ and the sigmoid $\delta(m')$ which is the same as Equation \ref{step1_BCE}.


\noindent\textbf{Attack Simulation for Adversarial Training.} Various attacks are common in practical image usage. Therefore, during the training process, we deploy an attack layer to watermarked images before employing a watermark extraction algorithm. This attack layer encompasses six common types of attacks: blur, Gaussian noise, brightness adjustment, contrast adjustment, saturation adjustment, and JPEG compression. To remain the differentiable of attack during training, we employ the differentiable simulation method to perform JPEG attack \cite{zhu2018hidden}.

\begin{table*}[]
\centering
\begin{tabular}{ccccccccc}
\hline

\multicolumn{1}{c}{\multirow{2}{*}{Methods}} & \multicolumn{6}{c}{Metrics}                                                                \\\hhline{|~|--------|}
\multicolumn{1}{c}{}                         & TPR(C.)$\uparrow$ & TPR(Adv.)$\uparrow$ & Bit Acc.(C.)$\uparrow$ & Bit Acc.(Adv.)$\uparrow $& FID$\downarrow $& CLIP $\uparrow$&PSNR$\uparrow$&SSIM$\downarrow$\\\hline
DwtDct        & 0.832 & 0.128  & 0.903   &  0.554   & 3.38 & 0.334 &39.2&0.974\\
DwtDctSvd     & 1.000 & 0.236  &  0.999   &  0.661   & 9.44 & 0.332 &39.0&0.982\\
RivaGan       & 1.000 & 0.714 & 0.999    & 0.829    & 15.3 & 0.333 &40.5&0.980\\
Tree-Ring     & 1.000 & 0.995 & --- & --- & 24.6 & 0.336 &---&---\\
Stable Signature &1.000  & 0.837 &  0.989   &     0.812& 13.4 & 0.335 &29.6&0.824\\
FSwatermark      & 1.000 & 0.914 &  0.999   &  0.872   & 21.7 & 0.334 &31.9&0.897\\
DistriMark (ours)  & 1.000 & \textbf{0.989} &  0.983   & \textbf{0.939}   & 14.6 &  0.334&30.8&0.856\\\hline
\end{tabular}
\caption{Watermark detection and traceability comparison. `C.' refers to results without image processing attacks. `Adv.' (Adversarial) refers to the average performance of a series of image processing attacks.}
\label{tab3:performance}
\end{table*}

\begin{table*}[htbp]
\centering
\begin{tabular}{*{13}{c}}
\hline
Scenario&Method& 1 & 2 & 3 & 4 & 5 & 6 & 7 & 8 & 9 & 10 & Comb.\\\hline 
\multirow{3}{*}{MLaaS}&DwtDct&0.587&0.929&0.497&0.479&0.598&0.512&0.493&0.498&0.524&0.494&0.507\\
&DwtDctSvd&0.622&0.999&0.634&0.675&0.997&0.516&0.506&0.554&0.634&0.512&0.512\\
&RivaGan&0.976&0.926&0.968&0.981&0.999&0.902&0.590&0.858&0.611&0.632&0.588\\\hline
\multirow{3}{*}{\begin{tabular}[c]{@{}c@{}}Model\\ Distrib.\end{tabular}}&{FSWatermark}&0.996&0.998&0.995&0.979&0.657&0.997&0.913&0.664&0.686&0.643&0.561\\
&Stable Signature&0.977&0.989&0.975&0.851&0.612&0.888&0.767&0.532&0.586&0.497&0.498\\\hhline{~------------}
&DistriMark (ours)&0.949&0.976&0.972&0.955&0.952&0.961&0.952&0.953&0.942&0.949&0.844\\
\hline
\end{tabular}
\caption{Bit accuracy of ten image processing attack. (1) Brightness, (2) Gauss Noise, (3) Contrast,  (4) Blur, (5) JPEG, (6) BM3D denoising algorithm, (7) Resize, (8-9) VAE-based compression algorithm and (10) diffusion-based 
reconstructive attack, respectively. `Comb.' indicates a mixture of the previous attacks.}
\label{tab:imageprocess}
\end{table*}

\section{Experiments}

\noindent\textbf{Implementation details.} In this paper, we focus on text-to-image generation, hence we utilized Stable Diffusion-v2 \cite{rombach2022high}. The number of inference steps is 25 for both generation and detection process. Following the settings of existing works \cite{wen2024tree}, we employ
the prompt from StableDiffsionDB \cite{wang2022diffusiondb} with guidance
scale of 5 during inference and an empty prompt during DDIM inversion. 
We utilize AdamW with a learning rate of $5 \times 10^{-4}$ and weight decay of 0.01 during finetuning. All experiments are conducted on a single NVIDIA L40.

\noindent\textbf{Watermarking baselines.} We select six typical baselines: three official watermark of Stable Diffusion \cite{rombach2022high} for cloud services called DwtDct \cite{cox2007digital}, DwtDctSvd \cite{cox2007digital}, and RivaGAN \cite{zhang2019robust}, two multi-bit watermarking methods named FSwatermark \cite{xiong2023flexible} and Stable Signature \cite{fernandez2023stable}, and a fine-tuning-free semantic watermarking method called Tree-Ring \cite{wen2024tree}.

\begin{figure*}[t]
\centering
\includegraphics[width=1.9\columnwidth]{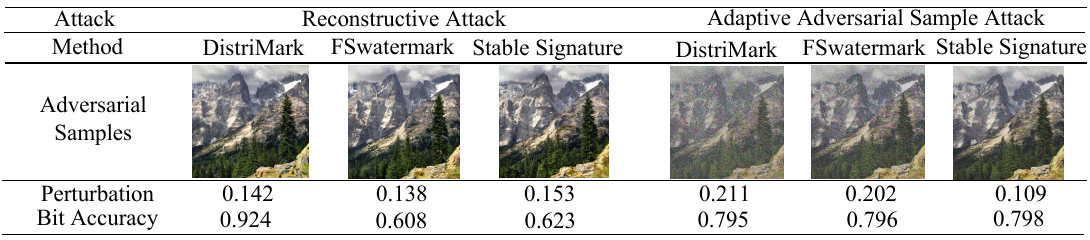} %
\caption{Adversarial samples obtained from adaptive adversarial sample attack and reconstrctive attack. }
\label{fig6}
\end{figure*}


\noindent\textbf{Evaluation metrics.} We measure the detection performance by the true positive rate (TPR) when the false
positive rate (FPR) is at 1\%. We measure the traceability performance by the bit accuracy. To measure the image generation quality, 
we compute the Peak Signalto-Noise Ratio (PSNR) \cite{hore2010image} and  Structural Similarity score (SSIM) \cite{wang2004image} for image distortion evaluation, the Fr$\acute{\text{e}}$chet Inception Distance (FID) \cite{heusel2017gans} and CLIP score \cite{Radford2021LearningTV} for image diversity and semantic evaluation, and Natural Image Quality Evaluator (NIQE) \cite{mittal2012making} and Perceptual Image Quality Evaluator (PIQE) \cite{venkatanath2015blind} for image quality evaluation.

\subsection{Watermark Security against Escape}

In order to make the watermark flexible for distributing the LDMs to a large number of model users with strong robustness, we leverage the semantic watermarking framework to inject the watermark message into the latent variable $m$. However, the model users can easily escape the watermark by replacing $m$ with other random latent variables to obtain the non-watermarked generated images. To tackle this issue, we design the security mechanism to decrease the generated image quality when the model user escape the watermark. Representative non-watermarked images under the security mechanism are shown in Figure \ref{fig8}. Embedding watermark does not significantly affect the image quality metrics NIQE and PIQE, or semantic quality metric CLIP. As the quality of unauthorized images decreases further, the quality of watermarked images also slightly deteriorates.

\subsection{Watermark Performance Comparison}
We compare the performance of our method with existing six typical baselines over two tasks: (1) \textbf{Detection}. We consider all compared methods as single-bit watermarks with a unified watermark. We set the FPR to be 1\% and test the TPR on 1, 000 watermarked images. (2) \textbf{Traceability}. Each compared method, excepting for the single-bit watermark Tree-Ring, serves as a multi-bit watermark. In our experiments, we assume that there are 1, 000 model users, each of them requires one watermark for model tracing. Each user generates 10 images, resulting in a dataset of 10, 000 watermarked images. During test, if an image contains a watermark, we then calculate the number of matched bits (Bit Accuracy) with the watermark of each user. The user with the highest Bit Accuracy is considered the traced user and verified. The comparison results are shown in Table \ref{tab3:performance}. Our watermarking achieves
strong robustness and significantly outperforms baselines in
both tasks. In terms of bit accuracy, it surpasses the
best-performing baseline by approximately 6.7\%. This can
be attributed to the extensive diffusion of the watermark
throughout the entire latent space, establishing a profound
binding between the watermark and the image semantics.



\subsection{Robustness against Image-Level Attacks}

We examine watermark's robustness against three typical kinds of adversarial attacks, including image processing attack \cite{song2010analysis} for transforming generated images, adaptive adversarial sample attacks \cite{jiang2023evading} for disturbing watermark verification, and reconstructive attacks \cite{balle2018variational,cheng2020learned,zhao2023invisible} for re-generating non-watermarked images. 


\noindent\textbf{Image Processing Attack.} We select ten representative types of image-level noise shown in Table \ref{tab:imageprocess}. Please refer to the Supplementary Materials for detailed parameter settings.

\begin{figure}[t]
  \centering
  \includegraphics[width=0.9\columnwidth]{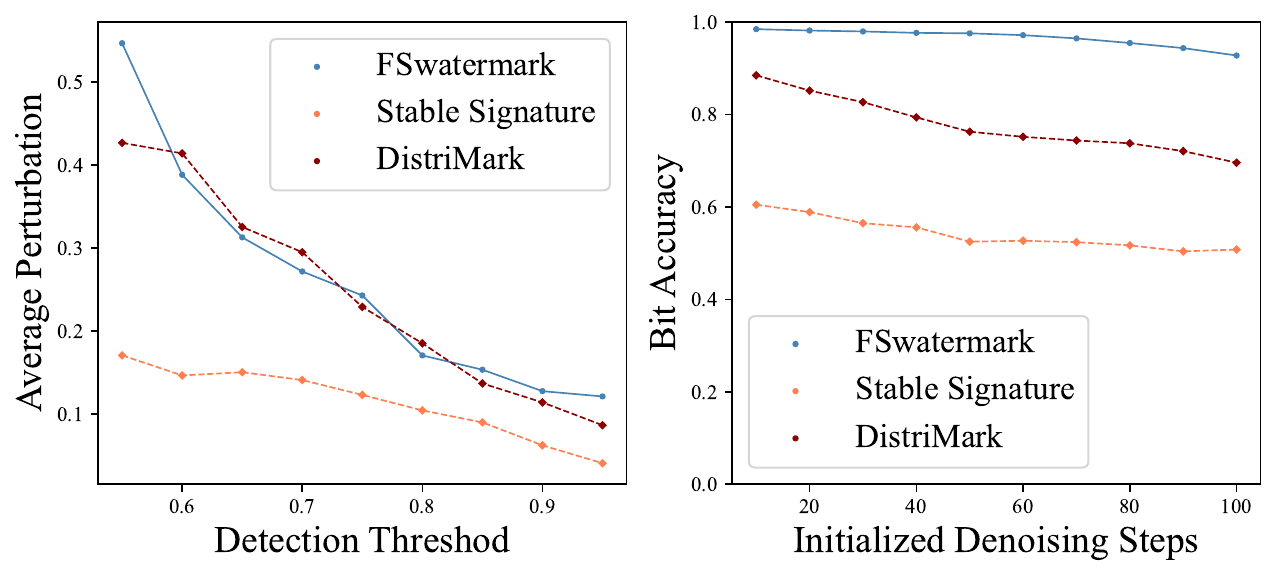}
  \caption{
  Results against adaptive adversarial sample attack (Left) and Reconstructive attack (Right).
  }
  \label{Fig:adaptive}
\end{figure}

\noindent\textbf{Adaptive Adversarial Sample Attack.}
To further enhance the attack ability, we assume that attackers can query a black-box watermark verification interface 
and conduct query-based black-box attack \cite{jiang2023evading}. 
By iterative querying the verification interface 
this attack compute optimal perturbations that progressively bring the watermark-free initial image closer to the original image.

\begin{figure}[t]
  \centering
  \includegraphics[width=0.9\columnwidth]{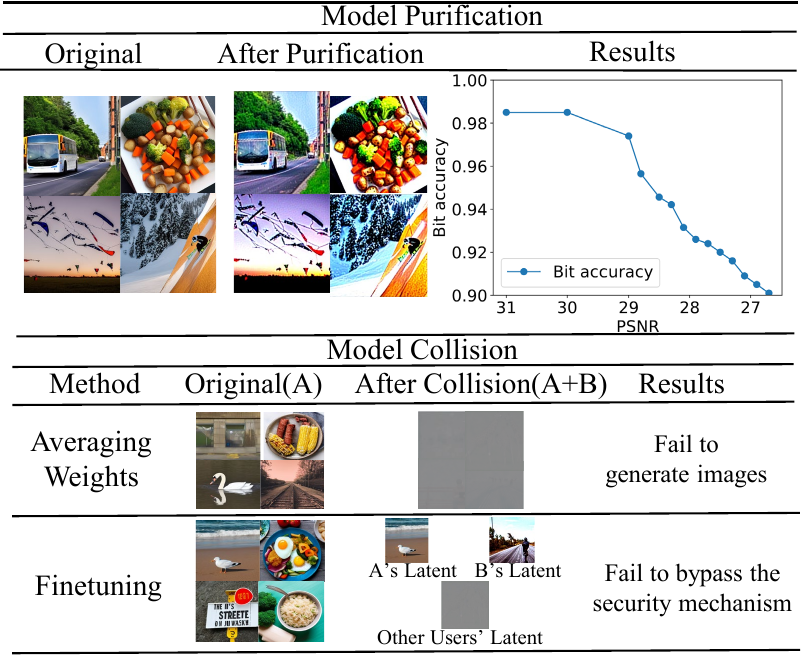}
  
  \caption{
 Results of model-level attacks.
  }
  \label{Fig:modelattack}
\end{figure}

\noindent\textbf{Reconstructive Attack.}
The core idea of the reconstructive attack is to add random noise to destroy the watermark and then reconstruct the image. We utilize the implementation of the paper \cite{zhao2023invisible} with denoising steps of 60.

\noindent\textbf{Main Results.}
 For each image processing attack, we report the average bit accuracy in Table \ref{tab:imageprocess}. We see that our DistriMark watermark is indeed robust  across all the transformations with the bit accuracy all above 0.9. DistriMark remarkably outperforms existing multi-bit watermarks on the average performance with more than 0.25 boost compared with the state-of-the-art (SoTA) results. Compared to the SoTA finetuning methods FSWatermark and Stable Signature, Distrimark has significant advantages in image resize, VAE-based compression algorithm, and reconstructive attack. Adaptive Samples Attack utilizes the same evaluation metric ${\ell}_{\infty}\mathrm{-norm}$ as described in the paper \cite{jiang2023evading}. Figure \ref{fig6} and Figure \ref{Fig:adaptive} display images and related parameters generated by Reconstructive Attack and Adaptive Adversarial Samples Attack under the same parameter conditions.  Under both types of attacks, DistriMark demonstrates better robustness than the other two methods. The robustness stems from the watermark being embedded at the semantic level within the image, so more extensive attacks are needed at the pixel level to remove the watermark. Please refer to the Supplementary Materials for more results.

\subsection{Robustness against Model-level Attacks}

Consider the model-level attacks of the model for both single users and multiple users, we have included two types of model-level attacks: model purification and model collusion. 

\noindent\textbf{Model purification.} The adversary fine-tunes the Variational autoencoder to circumvent watermark embedding through the same training mode as Section \ref{sec-3.2}. This involves removing the message loss $L_m$, and shifting the focus to the perceptual loss $L_w$ between the original image and the one reconstructed by the LDM auto-encoder. 


\begin{table}[]
\begin{tabular}{ccccccc}
\hline
{Method} & {PSNR} & {SSIM} & {FID} & {NIQE} & {PIQE} \\\hline
DwtDct            & 39.2 & 0.974 & 3.38  & 3.79 & 32.7 \\
DwtDctSvd         & 39.0 & 0.982 & 9.44  & 3.79 & 32.9 \\
RivaGan           & 40.5 & 0.980 & 15.3  & 3.82 & 32.4 \\
Tree-ring         & ---  & ---   & 25.9  & 4.25 & 33.5 \\
FSWatermark       & 31.9 & 0.897 & 21.7 & 4.22 & 34.7 \\
Stable Signature  & 29.6 & 0.864 & 13.4  & 3.79 & 33.7 \\
DistriMark& 30.8 & 0.856 & 14.6  & 3.98 & 34.2\\
\hline

\end{tabular}
\caption{Quality comparison of watermarked generated images.}
\label{tab2:quality}
\end{table}

\begin{table}[t]
\centering
\begin{tabular}{ccccc}
\hline
\multicolumn{1}{c}{\multirow{2}{*}{\begin{tabular}[c]{@{}c@{}}Skip 
\\ Connect\end{tabular}}} & {\multirow{2}{*}{\begin{tabular}[c]{@{}c@{}}Connect 
\\ Strenghth\end{tabular}}} & \multicolumn{2}{c}{Image Quality} & {\multirow{2}{*}{\begin{tabular}[c]{@{}c@{}}Bit 
\\ Acc.\end{tabular}}}\\\cline{3-4}
\multicolumn{1}{c}{}     && Watermark & Random \\\hline
\multicolumn{1}{c}{\multirow{3}{*}{Without}} & 0.1 &   31.7   &   30.3  & 0.986  \\
\multicolumn{1}{c}{}                         & 0.3 &   25.7   &   24.6  &  0.984 \\
\multicolumn{1}{c}{}                         & 0.5 &   24.1   &   23.7  &  0.979 \\\hline
\multirow{3}{*}{Single}                      & 0.1 &   29.6   &   24.5  &  0.981 \\
                                             & 0.3 &   25.8   &   17.2  & 0.979\\
                                             & 0.5 &   23.9   &   12.4  &   0.978\\\hline
\multirow{3}{*}{Multiple}                       & 0.1 &     30.8 &    14.7  &  0.985\\
                                             & 0.3 &     29.4 &   14.1   &  0.982\\
                                             & 0.5 &   26.3   &   12.1   &  0.982\\\hline

\end{tabular}
\caption{Evaluation on impact of skip connection.}
\label{tab4:skipConnection}
\end{table}

\begin{figure}[t]
  \centering
  \includegraphics[width=0.9\columnwidth]{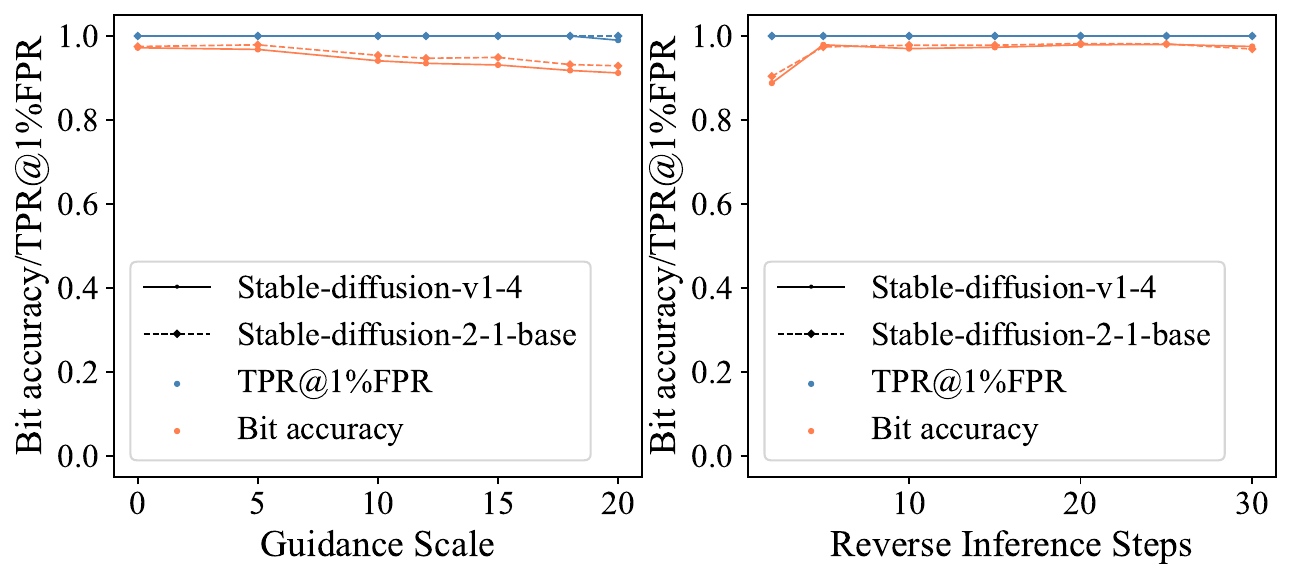}
  \caption{Ablation study results. (Left) Impact of guidance scales. (Right) Impact of reverse inference steps.  
  }
  \label{Fig:guidanceScale}
\end{figure}

\noindent\textbf{Model Collision.} We mainly considered two types of collusion attacks: Averaging weights and Finetuning. (1) Averaging weights. User$^{(i)}$ and User$^{(j)}$ can average their weights like Model Soup \cite{wortsman2022model} to creat a new model to deceive identification. 
(2) Finetuning. Another form of collusion attack is when the user B generates a large number of watermarked latent variables and watermarked generated images, and fine-tunes the VAE of A so that A can use B's watermark latent variables to generate images. 


\noindent\textbf{Main results.} Figure \ref{Fig:modelattack} shows the results of model purification attack. As for model purification, when the bit accuracy decreases, the image quality also declines. Empirically, it is difficult to significantly reduce the bit accuracy without affecting the image quality. As for model collision, because of the security mechanism, parameter averaging will cause a significant drop in image quality. This is because the watermark controller receives different watermark signals from different users, and directly performing model parameter averaging leads to a significant decline in image quality. As for finetuning, this could pose a threat of identity spoofing. However, it can be seen from the results, this still does not break the security mechanism. 

\subsection{Watermarked Image Quality}

Besides the  qualitative examples of how the watermarked  images are not sensitive for the human eyes to distinguish (see Figure \ref{fig8}), we further present quantitative evaluation of images generated by existing watermarking methods in Table \ref{tab2:quality}. The results show that no matter the qualitative metrics, our DistriMark
achieves comparable performance with existing works in the model distribution scenarios.  For DistriMark, the initial watermark is only manifested in the selection of the initial latent variables, hence the image quality is consistent with the diffusion model inference. 
Although DistriMark shows comparable quality compared to the methods in the model distribution scenario, the method involves a trade-off between the quality of unauthorized images and watermarked images when fine-tuning the VAE-Decoder, which results in lower image quality compared to post-processing methods.

\subsection{Ablation Study}



\noindent\textbf{Skip connection selection.}
The way latent variables are connected to the VAE-Decoder impacts how the VAE transforms images from the latent space to the pixel space. In Table \ref{tab4:skipConnection}, three methods were tested: no connection, single connection, and multi-level connection. Without skip connections, the model struggles to learn watermark characteristics, reducing the quality of images generated with watermarked latent variables. Multi-level connections improve feature learning and enhance image quality.

\noindent\textbf{Guidance scales.} Larger guidance scales result in more faithful of the generated image adherence to prompts. Following existing works \cite{wen2024tree}, we
cover the range of 0 to 20.  In
Figure \ref{Fig:guidanceScale} (left), although a higher guidance scale introduces errors in diffusion inversion due to the lack of such real guidance during detection, the watermark remains robust and reliable even at a guidance scale of 18.

\noindent\textbf{Number of the inversion step.} The inference step is often unknown in practice, which introduces a mismatch with
the inversion step. From Figure \ref{Fig:guidanceScale} (right) we can see that the number of inference steps does not significantly affect the accuracy of inversion which is beneficial in practice.

\section{Conclusion}
In this work, we propose a novel distribution scenario-oriented watermarking schema for diffusion models and a new security mechanism to prevent watermark leakage and watermark escape in the model distribution scenarios, which offers new insights into current distribution scenarios by considering watermark randomness and watermark-model association as key constraints for enhancing watermarking security. We separate the watermark injection from the security mechanism, ensuring that fine-tuning the VAE focuses solely on the security mechanism without the added task of learning watermark patterns. Our watermarking scheme ensures both security and efficiency in model distribution scenarios. In the future, our research directions will include adversarial methods against forge attack \cite{lukas2023leveraging}, security mechanism with higher image quality and more consideration about model distribution scenarios.

\bibliographystyle{named}
\bibliography{ijcai24}

\end{document}